\documentclass{article}

\usepackage{PRIMEarxiv}

\usepackage[numbers]{natbib}
\usepackage{amsmath}
\usepackage{amssymb} 
\usepackage[utf8]{inputenc} 
\usepackage[T1]{fontenc}    
\usepackage{hyperref}       
\usepackage{url}            
\usepackage{booktabs}       
\usepackage{amsfonts}       
\usepackage{nicefrac}       
\usepackage{microtype}      
\usepackage{lipsum}
\usepackage{fancyhdr}       
\usepackage{graphicx}       
\graphicspath{{media/}}     

\title{Geoinformation dependencies in geographic space and beyond 
\thanks{\textit{\underline{Citation}}: 
\textbf{Geoinformation dependencies in geographic space and beyond. Pages.... DOI:000000/11111. Contact: j.wang-4@utwente.nl.}} 
}

\author{
  Jon Wang \\
  Faculty of Geo-Information Science and Earth Observation (ITC), \\
  University of Twente, \\
  Hallenweg 8, 7522 NH, Enschede, The Netherlands \\
  \texttt{j.wang-4@utwente.nl} \\
   \And
  Meng Lu \\
  Department of Geography, \\
  University of Bayreuth, \\
  Universitaetsstrasse 30, Bayreuth, 95447 Germany\\
  \texttt{Meng.Lu@uni-bayreuth.de} \\
}

\begin{document}
\maketitle

\begin{abstract}
The use of geospatially dependent information, which has been stipulated as a law in geography, to model geographic patterns forms the cornerstone of geostatistics, and has been inherited in many data science based techniques as well, such as statistical learning algorithms. Still, we observe hesitations in interpreting geographic dependency scientifically as a property in geography, since interpretations of such dependency are subject to model choice with different hypotheses of trends and stationarity. Rather than questioning what can be considered as trends or why it is non-stationary, in this work, we share and consolidate a view that the properties of geographic dependency, being it trending or stationary, are essentially variations can be explained further by unobserved or unknown predictors, and not intrinsic to geographic space. Particularly, geoinformation dependency properties are in fact a projection of high dimensional feature space formed by all potential predictors into the lower dimension of geographic space, where geographic coordinates are equivalent to other predictors for modeling geographic patterns. This work brings together different aspects of geographic dependency, including similarity and heterogeneity, under a coherent framework, and aligns with the understanding of modeling in high dimensional feature space with different modeling concept including the classical geostatistics, Gaussian Process Regression and popular data science based spatial modeling techniques.
\end{abstract}


\section{Geographic dependency and its properties}

\begin{figure}[h]
\vspace{3mm}
 \begin{center}
 \includegraphics[width=1.\textwidth]{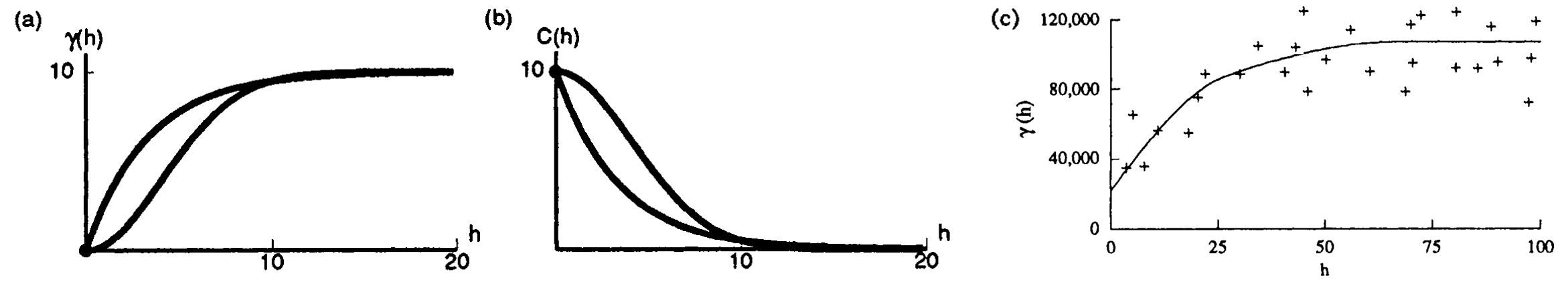}
 \caption{Classic variogram (a) and correlogram (b) examples (modified from \cite{isaaks1989geostats}), with \(\gamma\) the semivariance and \(c\) the covariance, and h the geospatial distance. Two variograms and their corresponding covariance functions are shown with different behaviors, especially at the origin. The one with a parabolic behavior near the origin is of the Gaussian model, while the other is of the exponential model. In application, the chosen model is fit to the observed samples (c).
 \label{fig:variograms}}
 \end{center}
\vspace{-8mm}
\end{figure}

When we say geographic pattern, one primary feature is that similar values of a variable of interest may be geospatially clustered or concentrated, rather than randomly scattered across geographic space. Geographic dependency describes the state of similar values of a variable observed at places near each other in geographic space \citep{getis2008history, griffith1992spatial, griffith2009spatial, goodchild2009problem}. Particularly, variable values are more similar at places that are closer-by. Since such dependency has been so frequently observed and can be statistically tested using metrics such as the \textit{Moran’s I} \citep{moran1947random, moran1948interpretation, cliff1973spatial}, it has been stipulated as a law in geography for the concept of spatial autocorrelation \citep{tobler1970computer, tobler2004first}. Apart from the \textit{Moran's I}, the dependency can be examined closer as a particular relationship between geographic proximity and similarity of variable values, or more often, it measures covariance as a function of geographic distance \citep{isaaks1989geostats}. One can use such a relationship to quantify the geographic pattern of a phenomenon and make predictions of the variable at geographic locations without observations. For example, with the technique of \textit{Kriging} modelling \citep{oliver1990kriging}, an unknown value as a weighted combination of the observed ones, where the weights are functions of the observation covariances only depends on locations, and the geographic dependency or spatial autocorrelation is explaining variations of phenomenon using geographic coordinates. When plotting geographic dependency as a function of geographic proximity, the semivariance is commonly adopted as a counterpart of the covariance \citep{bohling2005introduction}, hence, showing the difference of observations as a function of geographic distance, which can then be visualized in a variogram or correlogram (Figure \ref{fig:variograms}) \citep{isaaks1989geostats}. Parameters such as \textit{range}, \textit{sill} and \textit{nugget} have been introduced governing the shape of the function curve.

Being a law in geography, the dependency softens in scientific validity when complementary characteristics in geographic space were suggested, such as geographic heterogeneity \citep{goodchild2004validity}. Or, the definition of the law softens itself as stated “Everything is related to everything else, but (geographically) near things are more related than distant things”, where the word “near” and “more” does not restrict any exact intensity of dependency against nearness. Then what is more behind the geographic dependency to scientifically understand the similarity of observed values? Departing from Figure \ref{fig:variograms}, we can extend our understanding with at least two characters in the figure regarding how data samples are and are not explained by the variogram or correlogram.

\textbf{\textit{Basic character I: Stable geographic dependency that can be further explained}}

For the amount of information explained by the variogram, the similarity among the data samples is only a function of distance in geographic space, nothing else. How could this be? While this can be descriptively interesting enough for geographers as patterns are explained geographically, for those who would like to analytically know more about the mechanisms driving such patterns, there is always room for tracing back further about covariates \citep{rothman2019correlation}, or the drivers shaping the phenomenon. For instance, similarity of vegetation density quantified as the normalized difference vegetation index (NDVI) may be further due to the similarities, or dependency in other covariates, or driving factors such as elevation, precipitation, solar radiation, soil conditions or human activities \citep{tuoku2024impacts}, which underline fundamental biophysical processes, but are moderated or confounded by geographic locations. Mathematically, the geographical dependency, if quantified as covariance \(cov(z(\mathbf{s}_1), z(\mathbf{s}_2))=f(||\mathbf{s}||)\), in the values of the observed variable \(z\) at two geographic locations \(\mathbf{s}_1\) and \(\mathbf{s}_2\) (that can be in either two or three dimensions depends on space characterization), is essentially triggered by some other covariates that are also geographically dependent as \(cov(x(\mathbf{s}_1), x(\mathbf{s}_2))=f’(||\mathbf{s}||)\), with \(x\) being a covariate and \(f(||\cdot||)\) only a function of distance in the dimension of geographic space \(\mathbf{s}\). Depending on the explanation power of the covariate, one can decide to model the observations without geographic information \(\mathbf{s}\) as the covariate may explain all \citep{tuoku2024impacts}. Then geographic dependency seems to be manifesting those in other dimensions of covariates, as they also exhibit geographic patterns.  

It’s not always the case that observations and covariates exhibit stable geographic patterns as can be captured by variograms like those in Figure \ref{fig:variograms}. That is, hardly any phenomenon strictly meets the requirement of being, if not stationary, but second-order and intrinsically stationary \citep{myers1989or}.

\textbf{\textit{Basic character II: Instability or not is a question}}

There is also information not explained by the main curve. Similar values of a variable can still be observed when they are geographically far away from each other, whereas nearby values can be quite different, making the variogram “noisy”, as shown in Figure \ref{fig:variograms}. It means that the dependency exhibits unstable patterns in geographic space. Apparently, following the previous example of the NDVI, variables such as vegetation density are not exhibiting strict dependencies as a function of proximity in geographic space. In geostatistics, this has been referred to under different terminologies, such as heterogeneity, or more commonly, as the non-stationary dependency \citep{goodchild2009challenges}, meaning that the variance of the values and that of the difference in a pair of values separated by a fixed distance changes from place to place, and their covariance is no longer a function of just the distance, but also varies across locations.

Treating non-stationarity can be full of hesitation. Techniques such as trend removal, localized regression, or using non-stationary covariance functions are frequently in charge of modelling data with geographically non-stationarity \citep{vieira2010detrending, stewart1996geography, paciorek2003nonstationary, paciorek2006spatial, jun2008nonstationary}. One can either remove trend with the form of ordinary linear regression of observed variable \(z\) against either geographic locations \(\mathbf{s}\) as \(z=\theta \mathbf{s}\) before modelling again with stationary covariance, or include trends in non-stationary covariance functions in some more generic form as \(cov(z(\mathbf{s}_1), z(\mathbf{s}_2))=f(||\mathbf{s}||, \mathbf{s}_1, \mathbf{s}_2, c)\), which is not only a function of relative distance along \(\mathbf{s}\), but also depends on absolute locations on it, or also with an offset \(c\), such as \((s_n-c)\). In either way, it does not violate the fact there are other covariates underlining fundamental processes that are not completely manifested in geographic space. Hence, one can still model \(z=\theta'\mathbf{x}\) with \(\mathbf{x}\) as other covariates that can be found and also vary across geographic space with non-stationarity.

These two basic characters inspired several discussions about not only treating geographic heterogeneity as complements to the law of geographic dependency, but also using covariates as geographical contextual information of geographic locations when examining similarity \citep{anselin2020tobler, zhu2018spatial}. However, from our point of view, these discussions never clarify why there are mixed characters of being stationary or not. Even when it is non-stationary, the hesitation of choosing techniques such as trend removal or using non-stationary covariance functions reflect very different view towards the characters of geographic dependency under modelling, hence, bringing understanding of the geographic phenomenon largely subject to model choice, as opposed to exposing characters intrinsic to the phenomenon. Furthermore, other useful covariates, once found, may always bring extra dimensions of explanation power that modifies our understanding about dependency and similarities in geographic dimensions, because these covariates themselves may show patterns and thus be “collinear” with locational information, and moderate or confound the effect of geographic dependency during modelling. In its most general form, observed geographic patterns can be a mixture of hypotheses as shown in Expression (\ref{eq:generic}), where trend \(\mu\), the observations \(\mathbf{y}\) itself weighted by \(\mathbf{w}\), independent covariates \(\mathbf{x}\) as well as their neighboring ones \(\mathbf{x}_\gamma\) weighted by \(\mathbf{w}\) may all useful for the modelling of \(y\) (as adapted from \cite{elhorst2010applied, brunsdon1998spatial}). Hence, one has to decide how they should be mixed. These different options for model hypotheses continue to echo the earlier hesitation of what should be considered and modeled as an actual dependency pattern in geographic space \citep{myers1989or}.
\begin{equation}
y = \rho \mathbf{w} \mathbf{y} + \mathbf{x} \mathbf{\beta} + \mathbf{w} \mathbf{x}_\gamma + \mu.
\label{eq:generic}
\end{equation}

In this work, we will show how the basic characters listed above can be understood within one framework, where the observed variation in geographic space is indeed correlated to a mixture of factors including geographic locations as covariates. Modelling geographic patterns with these covariates either in terms of dependency along a covariate, such as dependency along geographic locations in the form of autocorrelation, or regressing against the covariates in the form of trend analysis is equivalent, but leads to very different interpretation! Hence, geographic autocorrelation, heterogeneity and similarity are just manifestations of higher dimensional variations in geographic space. At its ultimate implication, interpreting variations correlated to other dimensions of covariates exclusively as properties in geographic space, and even stipulating them as laws demands careful and further discussion.

The view above can be proved by:
\begin{itemize}
    \item The perspective of modelling geographic patterns explicitly with dependencies along the dimensions of covariates, which certainly include geographic locations as covariates among the others, is equivalent to the perspective of modelling with the Ordinary Linear (OL) regression along the covariates.
    \item Any unexplained variations along one dimension of the covariates should be taken up by other covariates as other dimensions. For instance, heterogeneity or non-stationarity presented in geographic space should be explained or modelled by some other covariates in either dependency- or OL regression-based manner. More progressively, if sufficient covariates are found, variations in the observed pattern should be stationary along all dimensions of covariates.
\end{itemize}

The good news is that the proof can be achieved by simply organizing and compiling existing theorems, which will be displayed in the following sections. The dependency-based approach is later represented as a more generalized Gaussian Process (GP) regression.

\section{Modelling geographic patterns with dependency and regression}
\label{sec:2}

In this section, we show that leveraging dependency characters, whether stationary or not in geographic space, is equivalent to applying OL regression by treating geographic locations as predictors or covariates. And the hesitation of choosing techniques between trend removal or modeling with non-stationary covariance function along geographic space are essentially hesitation in model choice, which leads to divergent interpretation and understanding. We also show that, as long as geographic locations can be treated equivalently as other covariates, modeling with dependency information in covariates other than geographic locations is an alternative to OL regression, leading to an extension of the concept of dependency into different dimensions of covariates. Ultimately, geographic characters of autocorrelation, heterogeneity and similarity in geographic space are representation of patterns nested in the high dimension of covariate space at lower dimension of geographic space, which are subject to modelling alternatives rather than independent “laws” of geography.

Mathematically, the proof of equivalence between using dependency information and OL regression has been paved more or less \citep{Rasmussen2006Gaussian, mackay1998introduction, MacKay2003}. If one attempts to model any observed variable \(z\) using any covariate \(x\) by following conventional OL regression form as:
\begin{equation}
p(z \mid \mathbf{x}, \mathbf{w}) \sim \mathcal{N}(f(\mathbf{x}), \sigma^2), \quad \text{where} \quad f(\mathbf{x}) = \boldsymbol{\phi}(\mathbf{x})^T \mathbf{w},
\label{eq:weight_space}
\end{equation}

where \(\sigma^2\) is the variance of gaussian noise \(\epsilon\), and \(\boldsymbol{\phi}(\mathbf{x})\) is a set of basis functions that project \(\mathbf{x}\) into high dimensional feature space if needed, such as \(x\mapsto \phi(x)=(1,x,x^2,x^3,\ldots)^T\), to involve nonlinearity for model flexibility \citep{Rasmussen2006Gaussian}. This formalization allows the structure in \(z\) to be fully specified by weight parameter \(\mathbf{w}\), hence, a parametric statistical modelling approach from the weight perspective. 

Within the \textit{Bayesian} framework, according to \cite{Rasmussen2006Gaussian}, the posterior estimation of the parameters \(\mathbf{w}\) can be achieved by treating Expression (\ref{eq:weight_space}) as the likelihood function, and involving a prior distribution of it with a covariance \(\Sigma_p\) such that \(\mathbf{w}\sim \mathcal{N}(0,\Sigma_p)\). Then the predictive distribution of any new predictor value \(\mathbf{x}_*\) is given by applying \(f\triangleq f(x)\) to \(\mathbf{x_*}\) as \(f_*\triangleq f(x_*)\) averaged over all possible weights encoded in the posterior distribution of W as:

\begin{equation}
\begin{split}
p(f_* \mid \mathbf{x}_*, \mathbf{x}, \mathbf{z}) & = \int p(f_* \mid \mathbf{x}_*, \mathbf{w}) p(\mathbf{w} \mid \mathbf{x}, \mathbf{z}) \, d\mathbf{w} \\
& \sim \mathcal{N} \Big( \boldsymbol{\phi}_*^T \Sigma_p \boldsymbol{\phi} (\boldsymbol{\phi}^T \Sigma_p \boldsymbol{\phi} + \sigma^2 I)^{-1} \mathbf{z}, \\
& \boldsymbol{\phi}_*^T \Sigma_p \boldsymbol{\phi}_* - \boldsymbol{\phi}_*^T \Sigma_p \boldsymbol{\phi} (\boldsymbol{\phi}^T \Sigma_p \boldsymbol{\phi} + \sigma^2 I)^{-1} \boldsymbol{\phi}^T \Sigma_p \boldsymbol{\phi}_* \Big),
\end{split}
\label{eq:weight_space_pred}
\end{equation}

where \(\boldsymbol{\phi}\triangleq\boldsymbol{\phi}(\mathbf{x})\) and \(\boldsymbol{\phi_*}\triangleq\boldsymbol{\phi_*}(\mathbf{x})\). Obviously, \(f_* \) is fully characterized by inner products, which are covariances, either within or across \(\mathbf{x}\)  and \(\mathbf{x_*}\) in the feature space \(\boldsymbol{\phi}(\cdot)\). And the explicit form of f becomes irrelevant for making predictions if the form of the inner product of any pair of predictors \(x\) and \(x'\) in the feature space can be specified, say as \(k\) in Expression (\ref{eq:weight_space_pred_k}). Then \(k=\boldsymbol{\phi}^T \Sigma_p \boldsymbol{\phi}\), \(k_{**}=\boldsymbol{\phi_*}^T \Sigma_p \boldsymbol{\phi_*}\) and \(k_*=\boldsymbol{\phi}^T \Sigma_p \boldsymbol{\phi_*}\). And Expression (\ref{eq:weight_space_pred}) can be rewritten as

\begin{equation}
p(f_* \mid x_*, X, Y) \sim \mathcal{N} \Big( k_*^T (k + \sigma^2 I)^{-1} Y, \, k_{**} - k_*^T (k + \sigma^2 I)^{-1} k_* \Big).
\label{eq:weight_space_pred_k}
\end{equation}

Specifying a covariance function is not new and has been known as “kernel tricks” in solving regression and classification problems by implicitly lifting predictors into high dimensional feature space \citep{mackay1998introduction, Rasmussen2006Gaussian}. This brings the other perspective of modelling \(f\) solely by mean \(m\) and covariance \(k\) as

\begin{equation}
f(\mathbf{x}) \sim \mathcal{GP} \big( m(\mathbf{x}), k(\mathbf{x}, \mathbf{x}') \big),
\label{eq:gp}
\end{equation}

where the symbol \(\mathcal{N}\) has been replaced by \(\mathcal{GP}\) for Gaussian Process emphasizing that the function values \(f\) together is considered as a random process with a mean function \(m\) over \(\mathbf{x}\) and a covariance function \(k\) specifying the joint distribution of the function values. In this way, the model is configured directly upon function values, hence, from a function perspective with no explicit function form being specified and the absence of parameters such as \(\mathbf{w}\) entitled the model as non-parametric.

The form of covariance function \(k\) can be specified in different ways depending on contexts \citep{Rasmussen2006Gaussian}, where we do need hyper-parameters to encode how \(f\) are jointly distributed across \(\mathbf{X}\). For instance, a frequently used covariance function is the Radial Basis Function (RBF) in the form of

\begin{equation}
k(\mathbf{x}, \mathbf{x}') = \sigma_f^2 \exp \left( -\frac{d(\mathbf{x}, \mathbf{x}')^2}{2l^2} \right) + \sigma^2 \delta,
\label{eq:kernel}
\end{equation}

where \(\sigma^2_f\), \(l\) and \(\delta\) are hyper-parameters used to describe how much, how far and how noisy the function values \(f\) is autocorrelated to itself across the feature space of predictor \(x\) disregarding any actual form of underlying \(f\). Optimizing hyper-parameters does not stipulate any fixed function form as \(\phi\) and \(\mathbf{w}\) do. This highlights the primary difference between the two perspectives as the interests in specifying \(k\) in the second perspective shifts away from the nontrivial specification of basis function \(\phi\) in the first perspective.

In fact, Expression (\ref{eq:gp}) and (\ref{eq:kernel}) encompass the concept of \textit{Kriging} in geostatistical modelling, but with a more generalized form that the mean and covariance functions can be of various forms and need to be estimated in parallel \citep{0e31d8b9596f4bbca24800b22c757ba9, christianson2023traditional}. Simply using geographic locational information such as the geographic coordinates \(\mathbf{s}=(s_{lon},s_{lat})\) to replace \(\mathbf{x}\), the covariance function in Expression (\ref{eq:kernel}) gives the sense of geographic dependency, while the mean function \(m\) in Expression (\ref{eq:gp}) gives deterministic patterns such as trends across geographic space. Obviously, the equivalence of the two perspectives, more specifically the weight perspective of OL regression and function perspective of GP regression, defines the equivalence in the roles of geographic locations and many other potential covariates in regression-based approaches. That is the geographic locations and other covariates are playing exactly the same role as predictors for different regression models. But the configurations of the models allow entirely different interpretations in different domains. In geospatial modelling, the GP regression allows the derivation of knowledge about geographic dependency. In a more generic situation with availability of multiple covariates, the two perspectives allows one to choose freely from them for modelling on each covariate, leading to a more general form of regression compared to Expression (\ref{eq:weight_space}) and (\ref{eq:gp}). For instance, when two covariates are used, the pattern of interest can be modelled as

\begin{equation}
f(\mathbf{x}_1, \mathbf{x}_2) = \boldsymbol{\phi}(\mathbf{x}_1)^T \mathbf{w} + g(\mathbf{x}_2), \quad 
g(\mathbf{x}_2) \sim \mathcal{GP} \big( m(\mathbf{x}_2), k(\mathbf{x}_2, \mathbf{x}_2') \big),
\label{eq:gp_mix}
\end{equation}

where \(f\) is explained by some predictor \(\mathbf{x_1}\) while autocorrelated with itself as a combination of existing noisy observations weighted by covariance function defined on \(\mathbf{x_2}\). While Expression (\ref{eq:gp_mix}) is in similar form of many other models, such as the mixed effects model with the GP part the random effect \citep{khan2023re}, it is moving towards the form of Expression (\ref{eq:generic}), where the two perspectives are also preeminently used! Simply replace \(\mathbf{x_2}\) with geographic locations \(\mathbf{s}\), it is straight-forward that the classic spatial model of \textit{Spatial Autoregressive} (\textit{SAR}) is a special case of it \citep{elhorst2010applied}. In the context of geostatistical modelling, one would hardly draw equivalency between the two terms on the right-hand side if \(\mathbf{x_1}\) and \(\mathbf{x_2}\) are assigned with different practical meanings. However, with the support of the two perspectives, it is now apparent that both \(\mathbf{x_1}\) and \(\mathbf{x_2}\) are predictors used for explaining \(z\).

\begin{figure}[h]
\vspace{3mm}
 \begin{center}
 \includegraphics[width=.75\textwidth]{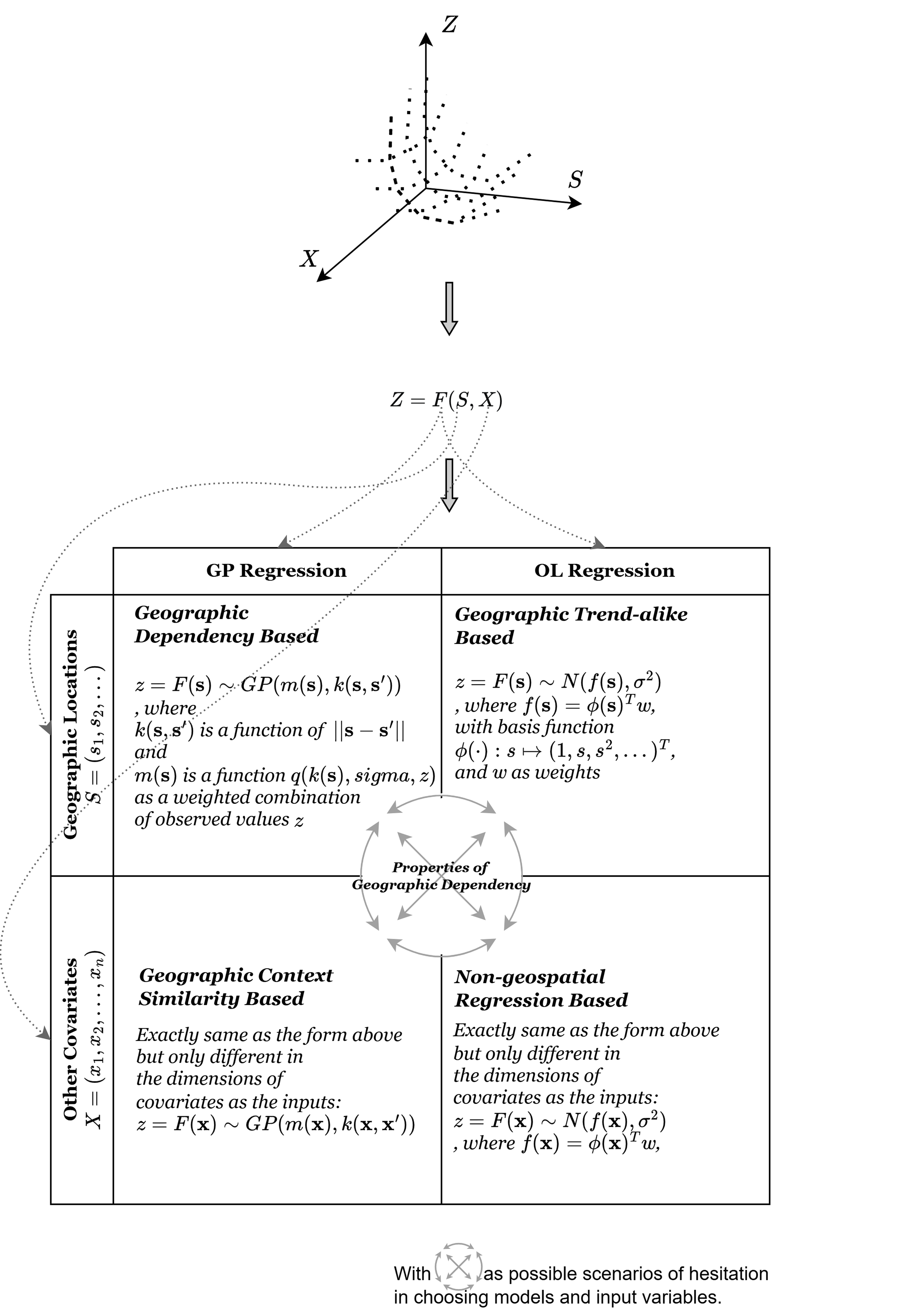}
 \caption{Framing together the options of using GP- and OL-based regressions, as well as geographic locations and other covariates as predictors for modelling geographic patterns with geographic dependency. The geographic dependency as a property is essentially driven by the choice of regression model and predictors.
 \label{fig:frame}}
 \end{center}
\vspace{-8mm}
\end{figure}

After replacing \(\mathbf{x_2}\) with geographic locations \(\mathbf{s}\) in Expression (\ref{eq:gp_mix}) as to model geographic pattern in the form of \(\mathbf{Z}=F(\mathbf{X},\mathbf{S})\), where notations are in capital bold indicating two types of spaces or two sets of vectors of geographic locations and other covariates. We can now officially put \(\mathbf{X}\) and \(\mathbf{S}\) side-by-side as equivalent covariates to be fed into different models according to the two perspective. By simplifing the dimensions of covariates into two major classes of spatial covariates and non-spatial ones, where the spatial covariates are of course the geographic locations, the observations \(\mathbf{Z}\) in geographic space that are potentially explainable by both geographic locations and other covariates can be visualized as samples in a pseudo three-dimensional space, similar to scatter plots as shown at the top of Figure \ref{fig:frame}, with geographic locations denoted as \(\mathbf{S}\) and non-spatial covariates as \(\mathbf{X}\). Both notations are in bold capitals since they present variables. For actual observations, \(\mathbf{S}\) can be a matrix of vectors of size either 2 or 3 depending on the involvement of latitude information, whereas \(\mathbf{X}\) can be vectors even longer with several possible covariates.

The two sets of equivalent elements, which are (1) geographic locations \(\mathbf{S}\) and other covariates \(\mathbf{X}\), and (2) two perspectives of regressions with GP- and OL-based, forming a 2x2 matrix of options for applying different regression models to different covariates (lower Figure \ref{fig:frame}). Across the columns, it is basically the issue of model choice regarding different perspectives of regression. According to Expression (\ref{eq:weight_space_pred_k}), choosing different covariance function is equivalent to stipulating feature space \(\phi(\mathbf{x})\) implicitly, hence, the model complexity, as jumping between column 1 and 2. Whereas jumping between row 1 and 2 is equivalent to choosing covariates given the regression choice or model complexity. It is rather simple scenarios in each entry of the matrix, which are illustrated already in previous sections. For instance, in the entry of (row 1, column 1) of the matrix, applying GP regression to geographic coordinates \(\mathbf{S}\) picks out the variations in the observation \(\mathbf{Z}\) that can be explained by dependency along covariates \(\mathbf{S}\), that is geographic dependency apart from some deterministic pattern modelled as \(m(\mathbf{x})\). (row 1, column 2) applies OL regression across geographic coordinates for modelling geographic patterns. While it takes the form of linear regression, the covariates, or geographic coordinates in this case, can be projected into high dimensional feature space (Expression (\ref{eq:weight_space})) for non-linearity and flexibility. The entry (row 2, column 1) applies GP modelling to covariates other than geographic locations, which is consistent with the more recent suggestion of modelling with similarity in geographic context variables \citep{zhu2018spatial, zhu2022third}. Although any covariate does come along with a geographic location, the similarity among them is essentially modelled in their own covariate space or dimensions, thus they would better not be referred to as geographical. In (row 2, column 2), probably it is the most regular approach of using potential covariates, such as that in the Land Use Regression model (LUR), disregarding geographic locations as predictors but may involve values of covariates within a buffer at each location \citep{ryan2007review, hoek2008review}.

While it is always flexible to switch among and combine the modelling options within the matrix, one may also encounter hesitations in choosing from the options, as both model choice and covariates selection can be arbitrary due to limited prior knowledge regarding model form and appropriate predictors. These selections also link to the interpretation of modelling rationale in the context of understanding geographic patterns. For instance, looking at (row 1, ) of the matrix, the choice of stationarity or not in column 1 may influence the modelling of the deterministic part \(m(\mathbf{s})\), while OL based regression in column 2 also overlaps with the \(m(\mathbf{s})\) in column 1. The hesitation in determining model forms would lead to significant differences in interpreting the geographic nature of the target variable. Apart from choosing modelling approaches, hesitation may also be triggered by the nature of the covariates. What if collinearity is found among them? Or, more specifically in the context of geographic patterns, what if some other covariates show patterns across geographic space producing collinearity between geographic locations and these covariates? Then one has to be careful with combining information across \(\mathbf{X}\) and \(\mathbf{S}\) with different models, that is, hesitation exists across both rows and columns as highlighted in the centre of the matrix in Figure \ref{fig:frame}.

Consequently, it is very unlikely that arbitrary combination of options in the table can lead to a perfectly complete explanation of variations in the observed patterns. As long as it is essentially regression models being applied to chosen covariates, the most general problem of incomplete or unknown covariates always exists. That is the consequence of missing covariate as one dimension would be always displayed as unexplained variations in other dimensions given some model configuration. Then what is left as unexplained variations would be projected onto each dimension of covariates as “noises”, and these “noises” lead to different interpretations. Looking back at the top part of Figure \ref{fig:frame}, all unexplained variations from other covariates and geographic locations, once projected onto the \(\mathbf{Z}-\mathbf{S}\) plane as “noises”, might be those explained as non-stationary. But this needs to be further proven.

\section{The dependency in geographic space and other dimensions}
\label{sec:3}

One intuition derived from the previous section is that the non-stationarity along the dimension of one covariate, such as geographic location, can be those unexplained variations from other dimensions of covariates. This can be illustrated as an extended proof. In fact, a much stronger conclusion has been reached and shown as a mathematical theorem that any random process is a sample of a stationary process in high dimension \citep{perrin2007can}. The theorem is stated as:

\textbf{\textit{Theorem I.}} Given any real-valued covariance matrix (hence, symmetric and positive definite with identical values on the diagonal) \(\mathbf{C}=(c_{ij})\) with \(i,j=1,2,\dots,s\), a real-valued positive definite function \(C\) on \(R^{(s+d)}\), where \(s+d>2\), and points \(z\in R^{(s+d)}\) exist, such that \(c_{ij}=C(z_{[si,di]},z_{[sj,dj]})\) with \(i,j=1,2,\dots,s\) \citep{perrin2003nonstationarity, perrin2007can, bornn2012modeling}.

Linking back to Figure \ref{fig:frame}, this conclusion is strong in the sense that only by using the GP regression-based approach shown as scenario in column 1, any stochastic process can be taken specifically as stationary in higher-dimensional space if higher dimensions of covariates can be found. This is fundamental as non-stationarity, which has been treated as a geographic property, is essentially a compressed summary of unexplained variations by only using existing covariates including the geographic locational information. The intuition was also given in the original work from \cite{perrin2007can} that with the extra dimensions of covariates properly specified, there is alway a hyperplane \(H\) exists, acting as a trend removal fit in the high dimensional covariate space, so that the original non-stationary function in low dimension of geographic space exhibits stationarity in high dimensional space. One can simply imagine that if techniques such as trend removal on the \(\mathbf{Z}-\mathbf{S}\) plane does not produce stationarity, there is always an opportunity to seek trend removal with extra dimensions of \(\mathbf{X}\) so that stationarity will display.

\begin{figure}[h]
\vspace{3mm}
 \begin{center}
 \includegraphics[width=1.0\textwidth]{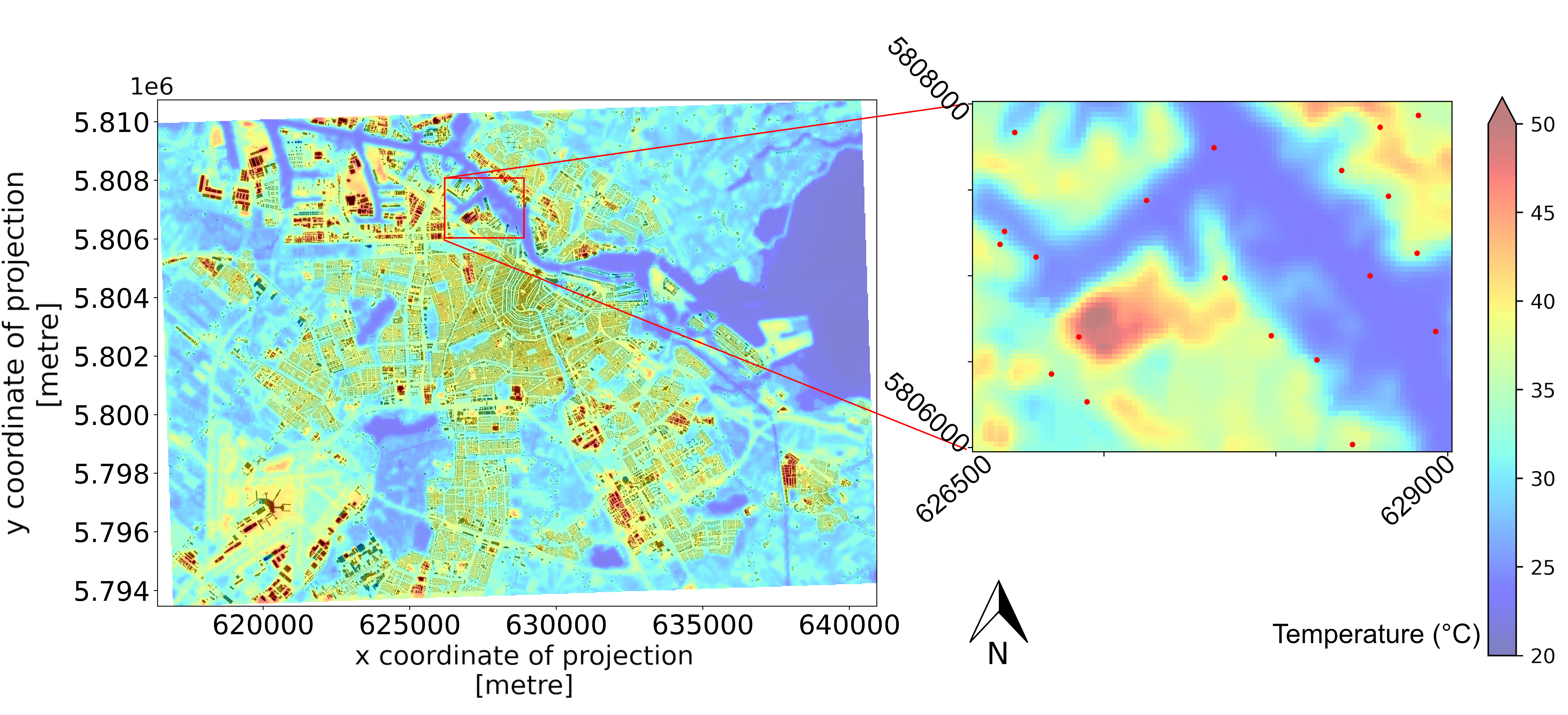}
 \caption{The land surface temperature (LST) of city of Amsterdam in the Netherlands is used as the study case, where a small fraction of the urban area containing different land cover types of built-ups, green and water bodies is chosen (left). Point samples are randomly chosen and shown as red dots on the image (right).  
 \label{fig:studyarea} }
 \end{center}
\vspace{-8mm}
\end{figure}

To further demonstrate the theorem and consolidate the intuition about the essence of geographic dependency, we follow the “\textit{dimension expansion}” approach \citep{bornn2012modeling}, which helps to show why non-stationarity is a compressed representation of high-dimensional patterns projected onto a lower dimension of geographic space. Imagine we have point sample observations of a land surface temperature (LST) dataset acquired from the satellite Thermal Infrared Sensor (TIRS) onboard Landsat 8 on July, 18th, 2021, of path 198, row 24, at local time 10:33 in the form of the level 2 (L2) product for the city of Amsterdam, the Netherlands \citep{landsat2022landsat}, where the samples are created as random draw from a small spatial region within the city. As can be seen in Figure \ref{fig:studyarea}, 20 point samples are drawn from either densely or sparsely built surfaces, or water bodies.

\begin{figure}[h]
\vspace{3mm}
 \begin{center}
 \includegraphics[width=.5\textwidth]{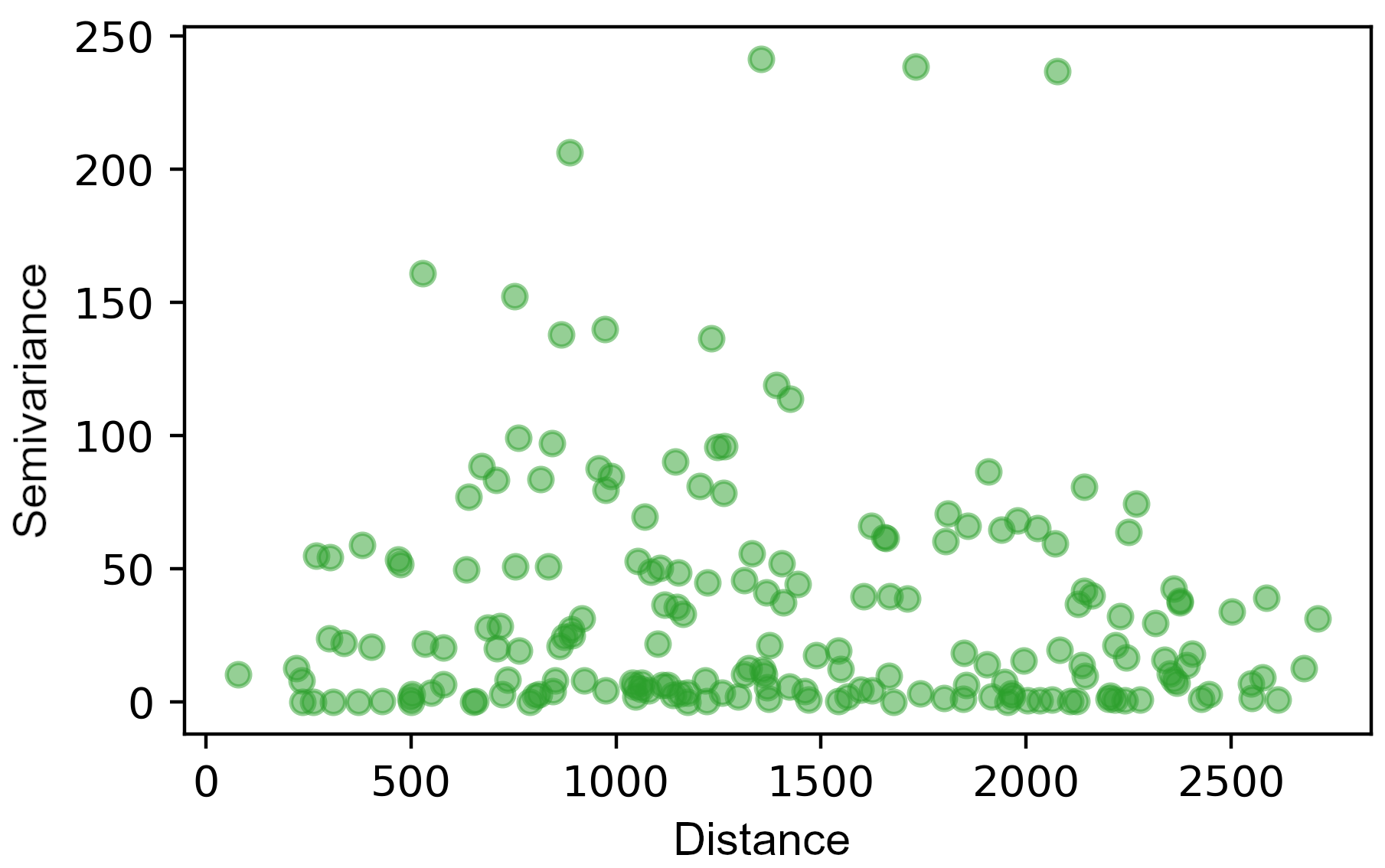}
 \caption{Semivariance scatterplot of the sampled LST values as a function of geographic distance.  
 \label{fig:variogram}}
 \end{center}
\vspace{-8mm}
\end{figure}

As the geospatial patterns of the LST are largely driven by the land surface material specifications \citep{li2013satellite, schwarz2011exploring}, one can already expect a non-stationary pattern in the function that can be fit into the samples as the samples are from contrasting land surface types. The non-stationarity, although not a property of the data sample, can be manifested by the variogram of the data samples as a function of the geographic distances among them (Figure \ref{fig:variogram}).

According to \textbf{\textit{Theorem I}}, a stationary function can be found and fit onto the data in a higher dimensional space formed by geographic locations and other properly specified covariates. Specifically, the word “properly” means that the extra covariates identified would not only explain more variations in the data, but also ensure that patterns left would be modelled as stationary. Consequently, in such newly constructed high-dimensional space, the function curve in the variogram should exhibit stationarity. As shown in Figure \ref{fig:frame}, this automatically links to the choice of covariates, which can be challenging without prior knowledge. The “dimension expansion” approach facilitates a non-parametric learning of unknown covariates as extra dimensions. Mathematically, it directly models the “locations” of the data samples in new dimensions introduced as extra covariates, so that the variogram measured in the higher-dimensional space converges towards a stationary pattern. Hence, let \(d_{ij}(\mathbf{x},\mathbf{s})\) be the distances among the data samples in the higher-dimensional space formed by geographic locations \(\mathbf{s}\) and values of newly added covariates \(\mathbf{x}\), and \(\gamma_{\phi}\) the variogram function fits into the space with extended dimension \(\mathbf{x}\), then \(\gamma_{\phi}(d_{ij}(\mathbf{x},\mathbf{s}))\) be the variogram that is stationary. A proper \(\mathbf{x}\) should exist, so that this high-dimensional space stationary variogram should be able to capture the dispersion patterns \(v^*_{ij}=0.5*(z_{si}-z_{sj})^2\) in the original observed values, thus both \(\mathbf{x}\) as new locations and parameter \(\phi\) of \(\gamma\) can be learnt by:

\begin{equation}
\hat{\phi}, \mathbf{Z} = \arg\min_{\phi, \mathbf{Z}'} \sum_{i<j} \left( v_{i,j}^* - \gamma_{\phi} \big( d_{i,j}([\mathbf{X}, \mathbf{Z}']) \big) \right)^2,
\label{eq:objective_loss}
\end{equation}

where \(\gamma_\phi\) can be of different forms of variogram function, and we adopt the Gaussian variogram as one of the most commonly used stipulated as:

\begin{equation}
\gamma(h) = \gamma_0^2 \left( 1 - \exp \left( -\frac{3h^2}{a^2} \right) \right),
\label{eq:kernel_studycase}
\end{equation}

where parameter \(\gamma_0\), \(h\) and \(a\) control the shape of the function \citep{bohling2005introduction}. Although the values, or locations along the new covariates are directly learnt with out parameterization, the setup above in Expression (\ref{eq:objective_loss}) still indicates that the number of \(\mathbf{x}\) must be chosen arbitrarily. That is the number of new dimensions of covariates. Here, as a demonstration and without repeatedly experimenting with different choices of number of dimensions, we simply explore how one extra dimension works for modelling the geographic dependency and possible interpretations. In this way, the original data samples shown in Figure \ref{fig:studyarea} can be thought of as living in a four dimensional space with the new covariate \(\mathbf{x}\), the two dimensions of geographic coordinates \(s_{lat},s_{lon}\), and the dimension of their observed LST values \(\mathbf{x}\). Once the optimization, or minimization in Expression (\ref{eq:objective_loss}) returns the optimal \(\mathbf{x}\), we could be able to visualize the high dimensional variogram fit to the data as shown in Figure \ref{fig:highdim_variogram} below, where the sample data points converged towards the stationary variogram function stipulated in Expression (\ref{eq:kernel_studycase}). Hence, a visual proof of Figure \ref{fig:frame} and also \textbf{\textit{Theorem I}} is reached as well.
\begin{figure}[h]
\vspace{3mm}
 \begin{center}
 \includegraphics[width=.5\textwidth]{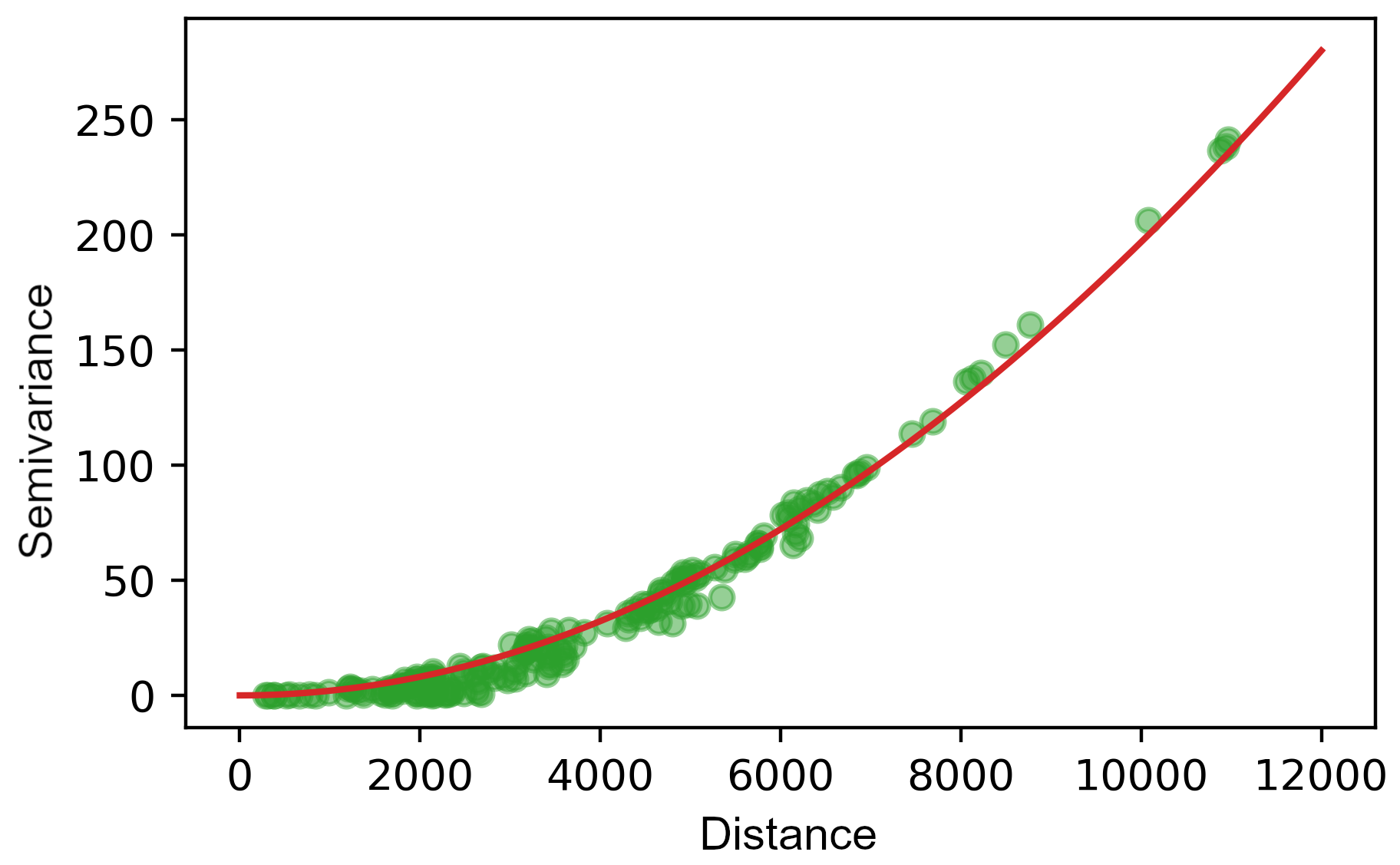}
 \caption{Semivariance scatterplot of the sampled LST values as a function of high dimensional distance extended by new covariate. The fit of a Gaussian variogram is superimposed.  
 \label{fig:highdim_variogram}}
 \end{center}
\vspace{-8mm}
\end{figure}

\begin{figure}[h]
\vspace{3mm}
 \begin{center}
 \includegraphics[width=1.0\textwidth]{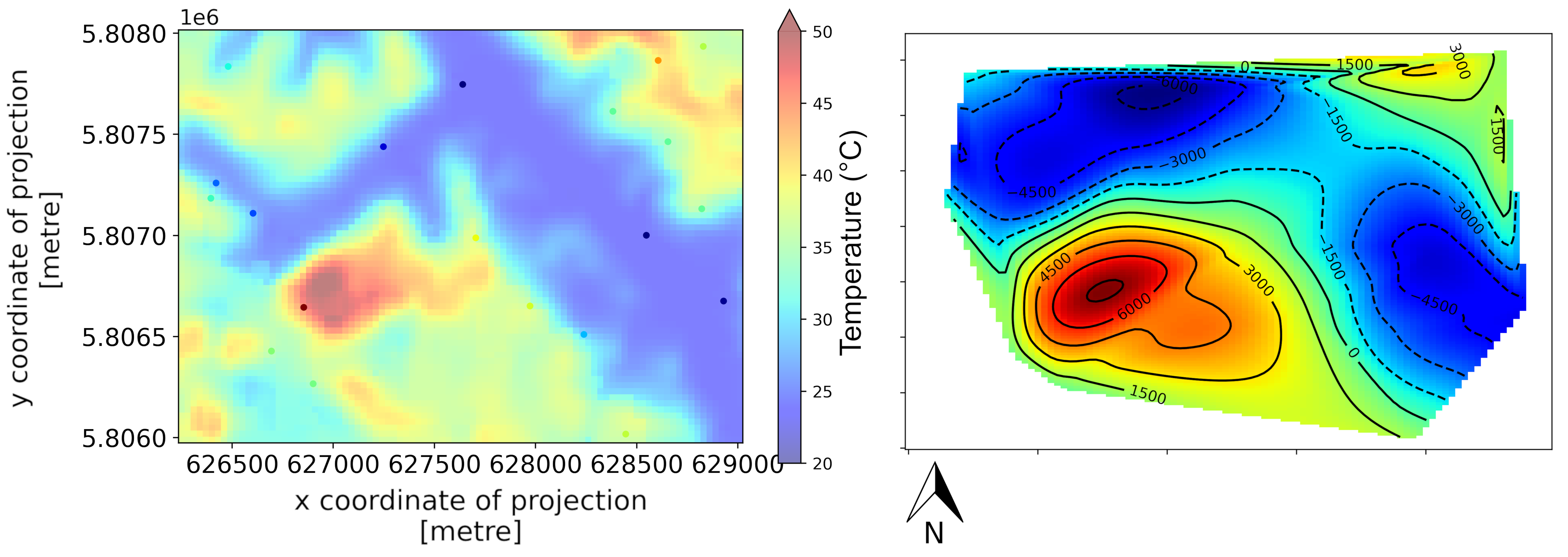}
 \caption{The learnt location values superimposed on the original LST map (left), and an interpolated location map of the new dimension within the extent of the samples is also shown with contour lines superimposed (right).  
 \label{fig:highdim_contour}}
 \end{center}
\vspace{-8mm}
\end{figure}

We can certainly explore further about the learnt dimension of covariate. By color-coding the data samples with the learnt values of the extra dimension (Figure \ref{fig:highdim_contour}), it can be seen that the values are resembling the contrasting temperature patterns between built surfaces and water-bodies, and also those across those built-up areas. While the values in the new dimension can also be considered as the “locations” along the dimension, this new dimension is primarily “pushing” contrasting measurements away from each other to divergent “elevations”, so that to be fit by a stationary variogram in high dimensional space. The contour lines provide a visualization of the “elevation” pattern in the new dimension. In this particular example with the LST pattern under investigation, the land surface specification is already a very strong governing factor of the temperature pattern, thus the “elevations” in the new dimension not only resembles the land surface specifications, but also the temperature values themselves! This becomes tricky in the sense that a good learnt dimension to explain the variations in the LST is of course an imaginary ideal covariate that resembles the LST itself, but this is also the consequence of choosing only one dimension to learn from Expression (\ref{eq:objective_loss}). However, this tricky situation turns out to be meaningful as it actually demonstrates the fact that once enough proper covariates are found, it can indeed be a stationary variogram function that fits into the sample data. And we can expect for now that if the number of new dimensions can be learnt more properly, they should be implying meaningful covariates that explain different parts of variations in the original samples. However, this is not the focus of this work, and has been briefly discussed in the work from \cite{bornn2012modeling}. Up to this point, we finish the demonstration of geographic dependency and the nature of its properties within and beyond the geographic space.

\section{Final remarks: Properties in geographic space and beyond}
\label{sec:4}

Everything, an entity, movement, event and phenomenon has its own location and span in geographic space and time. As geographers, we are very much interested in characterizing them by largely using locational information along geographic dimensions, being it in two or three dimensions. However, many factors driving these entities, movements, events and phenomena, while also being encompassed in geographic space, extend beyond the geographic space and exhibit patterns along their own dimensions. For instance, the age of the human population in a city may manifest minimal patterns across geographic space, but a simple histogram plot along the age values may exhibit clear patterns in the dimension of itself, and can be nicely approximated by basic statistical distributions. Any entity or phenomena largely driven by such factors, population in this case, exhibit patterns beyond the geographic space. Now, with the help of what has been framed in Figure \ref{fig:frame}, it is already clear that studying patterns in the geographic space may help us understand the target phenomenon, but can also compress patterns beyond the geographic space.

This issue of dimensions not only captures the conventional challenges in choosing predictors or covariates, and model assumption generally in all statistical based modelling setup, but also reflects the decade-long concern about defining trends and non-stationarity in the community of geostatistical modelling. For instance, “\textit{What is one person's nonstationarity (in mean) may be another person's random (correlated) variation…}” \citep{Cressie1986}, and “The weakly stationary with drift model also matches our predilection to assume that a deterministic part and a random part exist. In many applications, the deterministic part is viewed as signal and the random part as noise, the objective being to remove the noise. In the context of geostatistics, the random part is not viewed as noise and an obvious deterministic part may or may not be present… ” \citep{myers1989or}. Then a more fundamental question raised as: Whether non-stationarity, trends or even dependency should be considered as properties intrinsic to geographic space? Or is it just terminologies coined but lacking rigorous definitions?

It's undeniable that covariate choices and model assumptions involved in modelling specification leads to fundamental impacts on interpretation and understanding. Holding the hesitations in choosing covariates and models, as shown along the rows and columns in Figure \ref{fig:frame}, the hesitations propagate to knowledge derivation about geographic patterns. Thinking about a few scenarios as:
\begin{itemize}
    \item The observed variable exhibits stable, or stationary, dependency in geographic space, which can also be explained further by using other variables for extended understanding of the pattern.
    \item The observed variable exhibits non-stationary dependency in geographic space, but can still be modelled by using geographic locations, such as for modelling the trends. Again, other variables can be used as covariates to replace the locational information for the modelling. But whether it can still be viewed as a geographic pattern is in question, as geographic locations do not explain stable variations in the variable.
    \item The observed variable exhibits no pattern in geographic space, and has to be modelled by variables other than geographic locations.
\end{itemize}

where the interpretation can be driven into different directions. With real-world observations, normally, the observed variable needs to be modeled by a combination of two or more of the above scenarios, which is towards the most general form or tailored version of the geostatistical model represented as Expression (\ref{eq:generic}). In most cases, setting up models with mixed effects may be largely in the favor of modeling and prediction accuracy, which outweigh the purpose of understanding. Let alone with the recent enthusiasm in combining AI and machine learning components into geostatistical models \citep{appleby2020kriging, georganos2021geographical}. Modeling tricks such as the “dimension expansion” approach should be encouraged, albeit its infancy, it can indeed enhance understanding and thinking more about what properties are intrinsic and beyond geographic space, as the rationale behind geostatistical modelers.

\section*{Acknowledgments}

We are grateful to the open science community for all the tools such as the \textit{Python} modules and packages, and data such as the LandSat imagery that are freely available, leading to the possibility of all the experiments involved in this work.

\section*{Disclosure statement}

We declare no conflict of interest being involved in this work.

\section*{Notes on contributor(s)}

Wang and Lu conceptualize and implement the entire work collaboratively, including drafting the manuscript, implementing the model, coding, and editing the visuals.

\bibliographystyle{unsrt}  
\bibliography{geomanifold}

\begin{thebibliography}{10}

\bibitem{getis2008history}
Arthur Getis.
\newblock A history of the concept of spatial autocorrelation: A geographer's perspective.
\newblock {\em Geographical analysis}, 40(3):297--309, 2008.

\bibitem{griffith1992spatial}
Daniel~A Griffith.
\newblock What is spatial autocorrelation? reflections on the past 25 years of spatial statistics.
\newblock {\em L'Espace g{\'e}ographique}, pages 265--280, 1992.

\bibitem{griffith2009spatial}
Daniel~A Griffith.
\newblock Spatial autocorrelation.
\newblock {\em International encyclopedia of human geography}, 2009:308--316, 2009.

\bibitem{goodchild2009problem}
Michael~F Goodchild.
\newblock What problem? spatial autocorrelation and geographic information science.
\newblock {\em Geographical analysis}, 41(4):411--417, 2009.

\bibitem{moran1947random}
Patrick~AP Moran.
\newblock Random associations on a lattice.
\newblock In {\em Mathematical Proceedings of the Cambridge Philosophical Society}, volume~43, pages 321--328. Cambridge University Press, 1947.

\bibitem{moran1948interpretation}
Patrick~AP Moran.
\newblock The interpretation of statistical maps.
\newblock {\em Journal of the Royal Statistical Society. Series B (Methodological)}, 10(2):243--251, 1948.

\bibitem{cliff1973spatial}
Andrew~David Cliff and J~Keith Ord.
\newblock {\em Spatial autocorrelation}.
\newblock London, Pion, 1973.

\bibitem{tobler1970computer}
Waldo~R Tobler.
\newblock A computer movie simulating urban growth in the detroit region.
\newblock {\em Economic geography}, 46(sup1):234--240, 1970.

\bibitem{tobler2004first}
Waldo Tobler.
\newblock On the first law of geography: A reply.
\newblock {\em Annals of the association of American geographers}, 94(2):304--310, 2004.

\bibitem{isaaks1989geostats}
Edward~H. Isaaks and R.~Mohan Srivastava.
\newblock {\em Applied geostatistics}.
\newblock Oxford University Press, New York, 1989.

\bibitem{oliver1990kriging}
Margaret~A Oliver and Richard Webster.
\newblock Kriging: a method of interpolation for geographical information systems.
\newblock {\em International Journal of Geographical Information System}, 4(3):313--332, 1990.

\bibitem{bohling2005introduction}
Geoff Bohling.
\newblock Introduction to geostatistics and variogram analysis.
\newblock {\em Kansas geological survey}, 1(10):1--20, 2005.

\bibitem{goodchild2004validity}
Michael~F Goodchild.
\newblock The validity and usefulness of laws in geographic information science and geography.
\newblock {\em Annals of the Association of American Geographers}, 94(2):300--303, 2004.

\bibitem{rothman2019correlation}
Jessica Rothman, Monica~C Jackson, Kimberly~F Sellers, Talithia Williams, Subhash~R Lele, and Lance~A Waller.
\newblock Correlation induced by missing spatial covariates: a connection between variance components models and kriging.
\newblock {\em Journal of Mathematics and Statistical Science}, 5(12):333--344, 2019.

\bibitem{tuoku2024impacts}
Lina Tuoku, Zhijian Wu, and Baohui Men.
\newblock Impacts of climate factors and human activities on ndvi change in china.
\newblock {\em Ecological Informatics}, 81:102555, 2024.

\bibitem{myers1989or}
Donald~E Myers.
\newblock To be or not to be... stationary? that is the question.
\newblock {\em Mathematical Geology}, 21:347--362, 1989.

\bibitem{goodchild2009challenges}
Michael~F Goodchild.
\newblock Challenges in spatial analysis.
\newblock {\em The SAGE Handbook of spatial analysis}, pages 465--480, 2009.

\bibitem{vieira2010detrending}
Sidney~Rosa Vieira, Jos{\'e} Ruy Porto~de Carvalho, Marcos~Bacis Ceddia, and Antonio~Paz Gonz{\'a}lez.
\newblock Detrending non stationary data for geostatistical applications.
\newblock {\em Bragantia}, 69:01--08, 2010.

\bibitem{stewart1996geography}
A~Stewart~Fotheringham, Martin Charlton, and Chris Brunsdon.
\newblock The geography of parameter space: an investigation of spatial non-stationarity.
\newblock {\em International journal of geographical information systems}, 10(5):605--627, 1996.

\bibitem{paciorek2003nonstationary}
Christopher Paciorek and Mark Schervish.
\newblock Nonstationary covariance functions for gaussian process regression.
\newblock {\em Advances in neural information processing systems}, 16, 2003.

\bibitem{paciorek2006spatial}
Christopher~J Paciorek and Mark~J Schervish.
\newblock Spatial modelling using a new class of nonstationary covariance functions.
\newblock {\em Environmetrics: The official journal of the International Environmetrics Society}, 17(5):483--506, 2006.

\bibitem{jun2008nonstationary}
Mikyoung Jun and Michael~L Stein.
\newblock Nonstationary covariance models for global data.
\newblock 2008.

\bibitem{anselin2020tobler}
Luc Anselin and Xun Li.
\newblock Tobler’s law in a multivariate world.
\newblock {\em Geographical Analysis}, 52(4):494--510, 2020.

\bibitem{zhu2018spatial}
A-Xing Zhu, Guonian Lu, Jing Liu, Cheng-Zhi Qin, and Chenghu Zhou.
\newblock Spatial prediction based on third law of geography.
\newblock {\em Annals of GIS}, 24(4):225--240, 2018.

\bibitem{elhorst2010applied}
J~Paul Elhorst.
\newblock Applied spatial econometrics: raising the bar.
\newblock {\em Spatial economic analysis}, 5(1):9--28, 2010.

\bibitem{brunsdon1998spatial}
Chris Brunsdon, A~Stewart Fotheringham, and Martin Charlton.
\newblock Spatial nonstationarity and autoregressive models.
\newblock {\em Environment and Planning A}, 30(6):957--973, 1998.

\bibitem{Rasmussen2006Gaussian}
Carl~Edward Rasmussen and Christopher K.~I. Williams.
\newblock {\em Gaussian Processes for Machine Learning}.
\newblock The MIT Press, 2006.

\bibitem{mackay1998introduction}
David~JC MacKay et~al.
\newblock Introduction to gaussian processes.
\newblock {\em NATO ASI series F computer and systems sciences}, 168:133--166, 1998.

\bibitem{MacKay2003}
David J.~C. MacKay.
\newblock {\em Information Theory, Inference, and Learning Algorithms}.
\newblock Copyright Cambridge University Press, 2003.

\bibitem{0e31d8b9596f4bbca24800b22c757ba9}
J.P.C. Kleijnen.
\newblock Kriging: Methods and applications.
\newblock Workingpaper, CentER, Center for Economic Research, November 2017.

\bibitem{christianson2023traditional}
Ryan~B Christianson, Ryan~M Pollyea, and Robert~B Gramacy.
\newblock Traditional kriging versus modern gaussian processes for large-scale mining data.
\newblock {\em Statistical Analysis and Data Mining: The ASA Data Science Journal}, 16(5):488--506, 2023.

\bibitem{khan2023re}
Kori Khan and Candace Berrett.
\newblock Re-thinking spatial confounding in spatial linear mixed models.
\newblock {\em arXiv preprint arXiv:2301.05743}, 2023.

\bibitem{zhu2022third}
A-Xing Zhu and Matthew Turner.
\newblock How is the third law of geography different?
\newblock {\em Annals of GIS}, 28(1):57--67, 2022.

\bibitem{ryan2007review}
Patrick~H Ryan and Grace~K LeMasters.
\newblock A review of land-use regression models for characterizing intraurban air pollution exposure.
\newblock {\em Inhalation toxicology}, 19(sup1):127--133, 2007.

\bibitem{hoek2008review}
Gerard Hoek, Rob Beelen, Kees De~Hoogh, Danielle Vienneau, John Gulliver, Paul Fischer, and David Briggs.
\newblock A review of land-use regression models to assess spatial variation of outdoor air pollution.
\newblock {\em Atmospheric environment}, 42(33):7561--7578, 2008.

\bibitem{perrin2007can}
Olivier Perrin and Martin Schlather.
\newblock Can any multivariate gaussian vector be interpreted as a sample from a stationary random process?
\newblock {\em Statistics \& probability letters}, 77(9):881--884, 2007.

\bibitem{perrin2003nonstationarity}
Olivier Perrin and Wendy Meiring.
\newblock Nonstationarity in rn is second-order stationarity in r2n.
\newblock {\em Journal of applied probability}, 40(3):815--820, 2003.

\bibitem{bornn2012modeling}
Luke Bornn, Gavin Shaddick, and James~V Zidek.
\newblock Modeling nonstationary processes through dimension expansion.
\newblock {\em Journal of the American statistical association}, 107(497):281--289, 2012.

\bibitem{landsat2022landsat}
USGS Landsat.
\newblock Landsat 8-9 collection 2 (c2) level 2 science product (l2sp) guide.
\newblock {\em United States Geological Survey: Asheville, NC, USA}, pages 1--42, 2022.

\bibitem{li2013satellite}
Zhao-Liang Li, Bo-Hui Tang, Hua Wu, Huazhong Ren, Guangjian Yan, Zhengming Wan, Isabel~F Trigo, and Jos{\'e}~A Sobrino.
\newblock Satellite-derived land surface temperature: Current status and perspectives.
\newblock {\em Remote sensing of environment}, 131:14--37, 2013.

\bibitem{schwarz2011exploring}
Nina Schwarz, Sven Lautenbach, and Ralf Seppelt.
\newblock Exploring indicators for quantifying surface urban heat islands of european cities with modis land surface temperatures.
\newblock {\em Remote Sensing of Environment}, 115(12):3175--3186, 2011.

\bibitem{Cressie1986}
Noel Cressie.
\newblock Kriging nonstationary data.
\newblock {\em Journal of the American Statistical Association}, 81(395):625--634, 1986.

\bibitem{appleby2020kriging}
Gabriel Appleby, Linfeng Liu, and Li-Ping Liu.
\newblock Kriging convolutional networks.
\newblock In {\em Proceedings of the AAAI conference on artificial intelligence}, volume~34, pages 3187--3194, 2020.

\bibitem{georganos2021geographical}
Stefanos Georganos, Tais Grippa, Assane Niang~Gadiaga, Catherine Linard, Moritz Lennert, Sabine Vanhuysse, Nicholus Mboga, El{\'e}onore Wolff, and Stamatis Kalogirou.
\newblock Geographical random forests: a spatial extension of the random forest algorithm to address spatial heterogeneity in remote sensing and population modelling.
\newblock {\em Geocarto International}, 36(2):121--136, 2021.

\end{thebibliography}

\end{document}